\documentclass[conference,10pt]{IEEEtran}
\usepackage{times,amsmath,graphicx,flafter,indentfirst,float}
\usepackage{rotating}
\usepackage{graphicx}
\begin{document}
\title{Kinetics of Muller's Ratchet from Adaptive Landscape Viewpoint}
\author{\authorblockN{Shuyun Jiao\authorrefmark{1}\authorrefmark{2}, Yanbo Wang\authorrefmark{3}, Bo Yuan\authorrefmark{3}, Ping Ao\authorrefmark{1}\authorrefmark{4}}
\authorblockA{\authorrefmark{1}Shanghai Center for Systems Biomedicine, Key Laboratory of Systems Biomedicine of Ministry of Education,\\
 Shanghai Jiao Tong University, 200240, Shanghai, P.R.China}
\authorblockA{\authorrefmark{2}Department of Mathematics, Xinyang Normal University, 464000, Xinyang, Henan, P.R.China}
\authorblockA{\authorrefmark{3}Department of Computer Science and Engineering, Shanghai Jiao Tong University, 200240, Shanghai, P.R.China}
\authorblockA{\authorrefmark{4}Department of Physics, Shanghai Jiao Tong University, 200240, Shanghai, P.R.China\\
(Published by 2011 IEEE Conference on Systems Biology,  ISBN 978-1-4577-1666-9, pp: 27-32. Zhuhai, China, Sep 2-4)}
}

\maketitle
\begin{abstract}
Background: The accumulation of deleterious mutations of a
population directly contributes to the fate  as to how long the
population would exist.  Muller's ratchet provides a quantitative
framework to study the effect of accumulation.  Adaptive landscape
as a powerful concept in system biology provides a handle to
describe complex and rare biological events.  In this article we
study the evolutionary process of a population exposed to Muller's
ratchet from the new viewpoint of adaptive landscape which allows us
estimate the single click of the ratchet starting with an intuitive
understanding.

 Methods: We describe how Wright-Fisher process maps to Muller's ratchet. We analytically construct adaptive landscape from general diffusion equation.  It shows that the construction is  dynamical  and the adaptive landscape is independent of the existence and  normalization of the stationary distribution. We generalize the  application of diffusion model from adaptive landscape viewpoint.

Results: We develop a novel method to describe the dynamical
behavior of the population exposed to Muller's ratchet, and
analytically derive the decaying time of the fittest class of
populations as a mean first passage time.  Most importantly, we
describe the absorption phenomenon by adaptive landscape, where the
stationary distribution is non-normalizable.  These results suggest
the method may be used to understand the mechanism of populations
evolution and describe the biological processes quantitatively.
\end{abstract}

\begin{keywords}
 Wright-Fisher process, adaptive landscape, stationary distribution, mean first passage time
\end{keywords}

\section*{Background }
Muller's ratchet is a mechanism that has been suggested as an
explanation for the evolution of sex \cite{Maynard1978}.  For
asexually reproducing population, without recombination, chromosomes
are directly passed down to offsprings, as a consequence the
deleterious mutations accumulate so that the fittest class loses.
For sexually reproducing population, because of the existence of
recombination between parental genomes,  a parent carrying high
mutational loads can have offspring with fewer deleterious
mutations. The high cost of sexual reproduction is thus offset by
the benefits of inhibiting the ratchet \cite{Etheridge2009}. Muller
predicted ratchet mechanism in 1964.  Muller's ratchet has received
growing attention recently. Most studies of Muller's ratchet are
related to two issues. First is that without
recombination, the genetic uniformity of the offspring leads to much
lower genetic diversity, which is likely to make it more difficult
to adapt to \cite{Lampert2010}. Second is that population
lacking genetic repair should decay with time, due to successive
loss of the fittest individuals \cite{Maia2009}\cite{Barton2010}. In
addition, Muller's ratchet is relevant to some replicators,
endosymbionts and mitochondria. In order to assess the relevance of
Muller's ratchet, it is necessary to determine the rate (or the
time) for the accumulation of deleterious mutations
\cite{waxman2010}. It is widely recognized that the rate of
deleterious mutations being much higher than that of either reverse
or beneficial mutations can be a serious threat to the survival of
populations at the molecular level \cite{Maia2009}.
 It would have to be taken into account in any quantitative theory of the ratchet mechanism \cite{felsenstein1974}.\\
\indent Some biologists also have noticed a framework was needed to
construct such as \cite{felsenstein1974}\cite{waxman2010}. The
potential evolutionary importance of Muller's ratchet mechanism
would make it desirable to carry out careful quantitative studies on
its operation \cite{felsenstein1974}. There has been a continuous
progress in theory. The simplest and earliest mathematical model is
the pioneering work in \cite{Haigh1978}. The author In  \cite{Haigh1978} described the evolutionary process of asexual population of constant size on the condition of deterministic mutation selection
balance according to Wright-Fisher dynamics. And
the author found numerical evidence of relation between the total number of individuals and
the average time between clicks of the ratchet.  Authors in
\cite{stephan1993} treated pioneering model as diffusion
approximation, they  produced more accurate predictions over the
relatively slow regime. They noted the increasing importance of selection coefficient
for the rate of the ratchet for increasing values of the total number of individuals. The
author in \cite{Gessler1995} employed simulation and analytical
approaches to Muller's ratchet and estimated how different between
the distribution of mutations within a population and a Poisson
distribution.  The author in \cite{Etheridge2009} obtained diffusion
approximations for three different parameter regimes, depending on
the speed of the ratchet. The model mathematically embodied as stochastic
differential equation and shed new light on \cite{stephan1993}.
Authors in \cite{waxman2010} also mapped Muller's ratchet to
Wright-Fisher process, and got the prediction of the rate of
accumulation of deleterious mutations when parameters lie in the
fast and slow regimes of operation of the ratchet. In the present article,
inspired by recent works such as \cite{waxman2010}, we treat the
evolutionary process as  Wright-Fisher model for diffusion approximation,
analytically construct adaptive landscape, where the stationary
distribution is non-normalizable. This method expands the
application of the model in \cite{waxman2010}. The concept of
adaptive landscape is proposed by Sewall Wright to quantitatively
describe the complex biological phenomena as \cite{barton2011}.  Here adaptive landscape is clearly
quantified as potential function in \cite{Ao2005} from the physical
point of view.\\
\indent The key concept in constructing adaptive landscape is of potential function as a scalar function.
There is a long history of definition, interpretation,
and generalization of potential. Potential has also been applied to biological systems in various ways. The usefulness of a potential reemerges in the current
 study of dynamics of gene regulatory networks \cite{Ao2004A}, such as its application in genetic switch  \cite{Zhu2004}\cite{Zhu2004a}\cite{liang2010}. The role of potential is the  same as that of adaptive landscape. In this article, we do not distinct them.\\
\indent In this article we  present an approach to Muller's ratchet,
mapping a single click of Muller's ratchet to one dimensional
Wright-Fisher model where two allele classes are used to depict all different types of individuals in a population.
 The Wright-Fisher model  predicts the rate of  Muller's ratchet and suggests the accurate rate of  accumulation of  deleterious mutations could be derived more quickly by adaptive landscape. In addition we will demonstrate this method is independent of the existence and normalization of stationary distribution. Let we further address one dimensional  Wright-Fisher process to make our points more clearer.\\
\section*{Methods}
\subsection*{Model description}
 \indent We consider here an important and widely applied process- Muller's ratchet which is mapped to Wright-Fisher process. Muller's ratchet  is the process by which genomes of a finite  population composed of asexual individuals accumulate deleterious mutations in an irreversible manner \cite{muller1932}\cite{muller1964}. Consider a population of haploid asexual individuals with discrete
generations $t=0,1,2,\ldots$. The starting point in a generation is regarded as adult stage, after all selection has occurred and
immediately prior to reproduction. New mutations occur at
reproduction and all mutations are assumed to deleteriously affect
viability but have no effect on fertility. Supposed population size
is always fixed  each generation, then there are always  alleles
with constant number in the gene pool. Here we consider one locus with two alleles $A$ and $a$, that is, there are two classes in the haploid asexual population.
  One class is individuals with allele $A$ while the other is individuals with allele $a$. Supposed mutation from allele $A$ to $a$ is deleterious, we assume population of constant size $N>1$,  generations are
non-overlapping. The lifecycle of the individuals in the population
is from adults to juveniles, during which we consider the change of
allele $A$ in the presence of irreversible mutations, selection and
random  drift. The change of fittest allele frequency  is exposed
to Muller's ratchet. Given above assumption, we can study the
evolution of allele $A$. The frequency of  allele  $A$ in generation
$t$ is $x_{t}$ while that of allele $a$ is $1-x_{t}$, $x_{t}=
0,1/N,\ldots,1$. Let $\mu$ be the probability that an offspring of
an adult with allele $A$ is an individual with allele $a$. It is
labeled by $M_{1,0}$, that is, $M_{1,0}=\mu$. Analogously,
$M_{0,0}=1-\mu$, $M_{0,1}=0$, $M_{1,1}=1$. The relative viability of
individuals with allele $A$ is $\nu=1$ while that of individuals
with allele $a$ is $\nu_{1}=1-\sigma$, where $\sigma$ can be treated
as  effective selection rates associated with deleterious
mutations. Then in generation $(t+1)$, the frequency of allele $A$
is
\begin{equation}
x_{t+1}=\frac{(1-\mu)x_{t}}{1-\sigma+\sigma(1-\mu)x_{t}}.
\end{equation}
Under the general diffusion approximation, frequency $x_{t}$ is
treated as continuous quantities $x$, and this also leads to the
distribution of the frequency of  allele $A$ being probability
density. Let $\rho\equiv\rho(x,t)$ be the probability density of the
frequency of allele $A$ being $x$ at time $t$. The diffusion
equation obeys the following formula
\cite{Kimura1964}\cite{ewens1979}
\begin{equation}
\partial_{t}\rho=\partial_{x}^{2}[V(x)\rho/2] -\partial_{x}[M(x)\rho],
\end{equation}
with
\begin{eqnarray}
M(x)&=& \frac{x[(\sigma-\mu)-\sigma(1-\mu)x]}{1-\sigma+\sigma(1-\mu)x},\\
V(x)&=&\frac{x(1-x)}{N},
\end{eqnarray}
where $M(x)$ represents the average transition rate of $x$ or driving force \cite{Kimura1964}\cite{John1998} and $V(x)$ is the corresponding variance.
\subsection*{Adaptive landscape}
\indent We can also depict the same evolutionary process by the symmetric equation
\begin{equation}
\label{2}
\partial_{t} \rho = \partial_{x} [\epsilon D(x)\partial_{x}-f(x)] \rho
\end{equation}
with
 \begin{eqnarray}
\label{3}
    f(x) &=&\frac{x[(\sigma-\mu)-\sigma(1-\mu)x]}{1-\sigma+\sigma(1-\mu)x}-\frac{1-2x}{2N}, \\
    \epsilon D(x) &=&\frac{x(1-x)}{2N}.
\end{eqnarray}
Adaptive landscape is directly given by
\begin{equation}
\Phi(x)=\int \dfrac{f(x)}{D(x)}dx.
\end{equation}
We are interested in the dynamical property of adaptive landscape,  so we treat $\Phi$ and $\Phi/ \epsilon$  no different in this respect, that is, for convenience we can take $\epsilon=1$ of $\epsilon D(x)$.
\begin{eqnarray}
\label{7}
\Phi(x)&=&\int \dfrac{f(x)}{\epsilon D(x)}dx  \\
 &=&\frac{2N \mu(1-\sigma)}{  1-\sigma \mu}\ln(1-x)-\ln x(1-x)   \nonumber \\
\quad &&+\frac{2N(1-\mu)}{1-\sigma \mu}\ln(1-\sigma+x\sigma(1-\mu)).\nonumber
\end{eqnarray}
Here the adaptive landscape is composed of three terms. The first term and the third term quantify the effect of irreversible mutation and selection respectively, the second term quantifies the effect of random drift. \\
\indent In addition, the symmetric Eq.(5) has two advantages. On the one hand, the adaptive landscape is directly read out when the detailed balance is satisfied. On the other hand, the constructive method is dynamical, independent of existence and normalization of stationary distribution.  We call $f(x)$ directional transition rate, integrating the effects of $M(x)$ and the derivative of $V(x)$. Directional transition rate can give equilibrium states when $f(x)$ is linear form. \\
\indent Under the very natural boundary condition satisfying the
probability flux of the system is zero, we  assume the
probability flows in $x\in [0,1]$.
 The  stationary distribution for diffusion approximation is given by
\begin{eqnarray*}
\label{5}
\rho(x)= \dfrac{1}{Z}\exp \left(\dfrac{\Phi(x)}{\epsilon}\right).
\end{eqnarray*}
The stationary distribution can also be expressed as
\begin{eqnarray}
  \lefteqn{ \rho(x) \propto  \exp \left(\frac{\sigma \mu-1-2N \mu(\sigma-1)}{1-\sigma \mu}\ln(1-x)\right.} \nonumber \\
&& \quad \left .-\ln x + \frac{2N(1-\mu)}{1-\sigma \mu}\ln(1-\sigma+x\sigma(1-\mu))\right),  \nonumber \\
\end{eqnarray}
it has the form of Boltzmman-Gibbs distribution \cite{Ao2008A}, so the scalar function $\Phi(x)$ naturally acquires
 the meaning of potential energy \cite{Ao2004A}. The value of $Z$ determines the normalization of $\rho$ from the perspective of probability,
  and the finite value of $Z$ manifests the normalization of $\rho$. The constant $\epsilon$ holds the same position as temperature of Boltzmman-Gibbs distribution in statistical mechanics. But it does not hold the nature of temperature in Boltzmman-Gibbs distribution.
\section*{Results and Discussion}
Previous works mainly focus on the parameters ranges that either lie
at the lower mutation rates regime or the higher mutation rates
regime based on stochastic differential equations. We analyze the evolutionary
process  in all parameters regimes based on
Fokker-Planck equation by adaptive landscape. We investigate the
single click time analytically.
\subsection*{Irreversible mutation, selection and random  drift balance}
\indent To understand the mechanism of Muller's ratchet, a full characterization of dynamical process is a prerequisite for obtaining  more accurate decaying time. Here we study it by adaptive landscape in detail.\\
 \indent Concretely we divide the regime of mutation rates into three sections. Mutation rates $\mu \in (0,1/(2N-1)]$, selection rates $\sigma \in(\mu,2\mu/(1+\mu)]$, this condition corresponds to the case that selection rates and mutation rates lie in the lower regime. In the medium regime mutation rates $\mu \in(1/(2N-1),1/(2N-2)]$, selection rates $\sigma \in(\mu,1)$. In the regime of high mutation rates,  it is possible that the number of mutation is greater than one in the process. Concretely, mutation rates $\mu \in(1/(2N-2),1)$, selection rates $\sigma \in(\mu,1)$.
The adaptive landscape of full parameter regimes  is visualized as
Fig.1, From the expression and visualization of adaptive landscape
$\Phi(x)$, we may find there are two singular points $0$ and $1$ of
adaptive landscape, characterized by infinity value in Fig.1. In the
low and medium mutation rates regimes, singularity indicates the
population stable, but in the high mutation rates regime, singularity
from $x=0$ indicates the population stable while singularity from
$x=1$ means the population unstable.
\begin{figure}[h] \centering
\includegraphics[width=9cm]{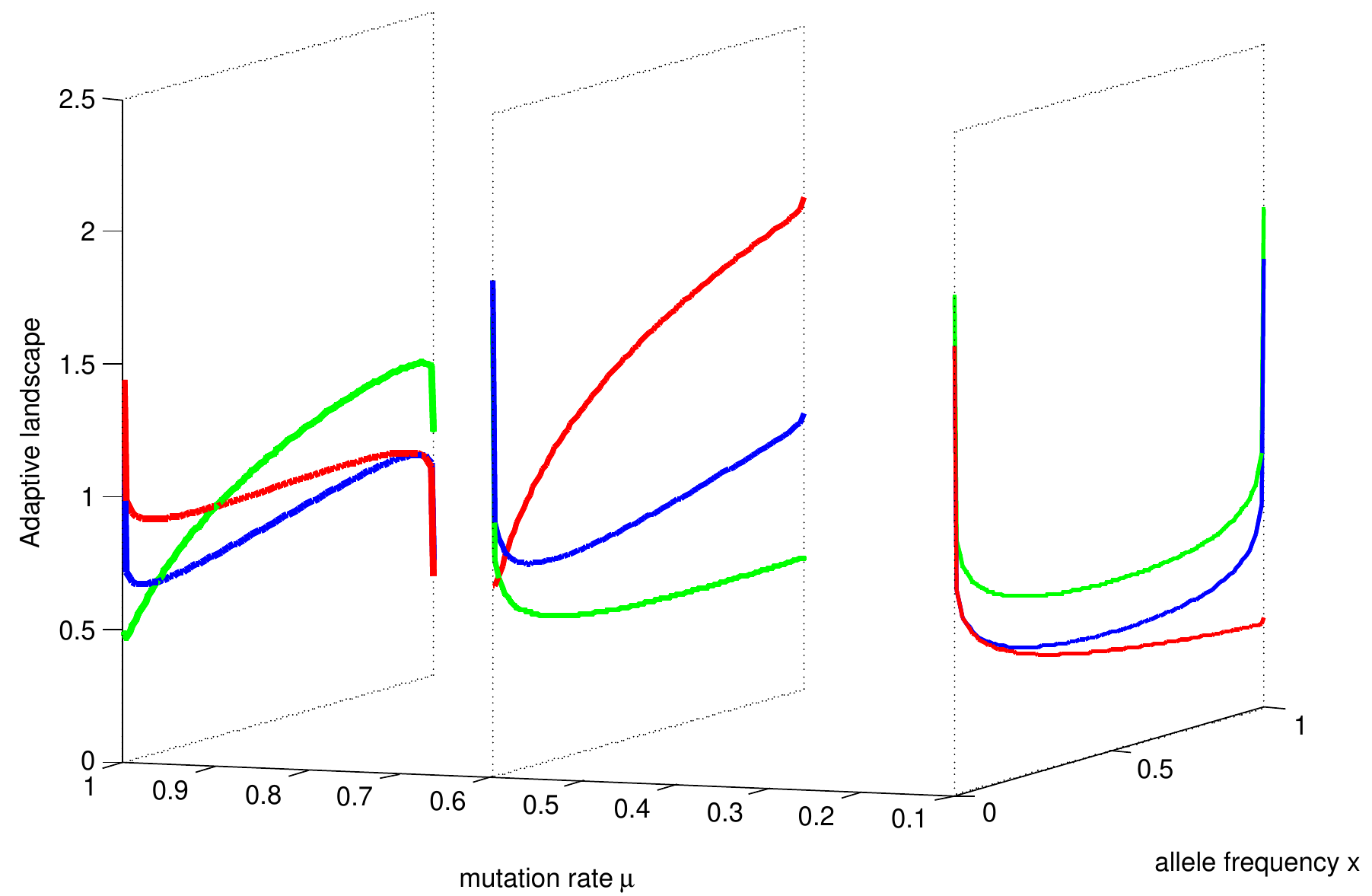}
\caption{\small \indent Adaptive landscape against allele frequency $x$ with mutation rates and selection rates in all regimes. $x$ label represents allele frequency, $\mu$ label represents mutation rates while vertical label is corresponding value of $\Phi$. Two stable states(maximums of adaptive landscape) occur in the low and high mutation rates regimes. Assume population size is constant $N$=50. In the lower regime, green  curve represents $\mu$=0.000005, $\sigma$=0.00005, blue curve stands for $\mu$=0.000005, $\sigma$=0.019, red curve represents  $\mu$=0.01,$\sigma$=0.019. In the medium mutation rates regime, green curve represents  $\mu$=0.01015, $\sigma$=0.015, blue curve stands for  $\mu$=0.01015, $\sigma$=0.05, red curve investigates  $\mu$=0.01015, $\sigma$=0.9. In the higher mutation rates regime, green curve represents $\mu$=0.4, $\sigma$=0.9, blue curve stands for  $\mu$=0.02, $\sigma$=0.1, red curve describes the case for  $\mu$=0.02, $\sigma$=0.05.}
\end{figure}
Fig.1 demonstrates the whole process of the population evolution
including the forming and losing the fittest class. With increasing
selection rates the fittest class $A$ is forming quickly while with
increasing mutation rates the fittest class $A$ loses easily. In the
lower mutation rates regime it describes the population is
likely to move to the fittest class with increasing selection rates,
the process is dominated by selection. Blue and red
curves manifest the losing process of allele $A$ with accumulation of
irreversible mutations. Because the mutation rates are lower, selection rates dependent of mutation rates
are lower, these factors result in the difficult change of
fittest class. As a consequence there are two fittest classes in the process. In the middle mutation rates regime, it demonstrates the transition process between lower and higher mutation
rates. The number of mutations is also fewer than one in one
generation in the process. In the
higher  mutation rates regime, green curve
describes the population is likely to move to the fittest class so
that the population exists in the form of coexistence of $A$ and $a$. Red and blue curves manifest the losing process  with accumulation of irreversible
mutations. Because mutation rates are
higher, as a consequence allele $a$ occurs to a certian percentage. Selection rates
dependent of mutation rates tend to survive allele $A$.  There are
two stable states in the process under the balance of irreversible mutation and selection. The evolutionary process is
dominated by the irreversible mutations, the fittest class $A$
loses. So we can draw the conclusion that the click process occurs
when there are two stable states in the process.
\subsection*{The single click time for Muller's ratchet}
\indent We visualize adaptive landscape, then one may wonder about how the population moves from one peak to another and how long it might be to move from one maximum to another.
 The process was first visualized by Wright in  1932. In addition, the problem of transition from metastable states is ubiquitous in almost all scientific areas.
  Most of  previous works  encounter  finite potential barriers from the physical point of view.
  Interesting issue here is that we touch upon  infinite potential barriers under the circumstance of  well defined two stable states. Then we manifest the derivation of a single click time. The time of a click of the ratchet is recognized as the random time of loss of the fittest class as \cite{waxman2010}. \\
\indent The single click time  is well defined when two fittest classes exist in the process. It means the interval time from the forming of one fittest class  to its extinction. The corresponding processes are that there are two well-defined stable states in the system.
 Corresponding graphs of adaptive landscape is the following:\\
\begin{figure}[h]
  \centering
\includegraphics[width=9cm]{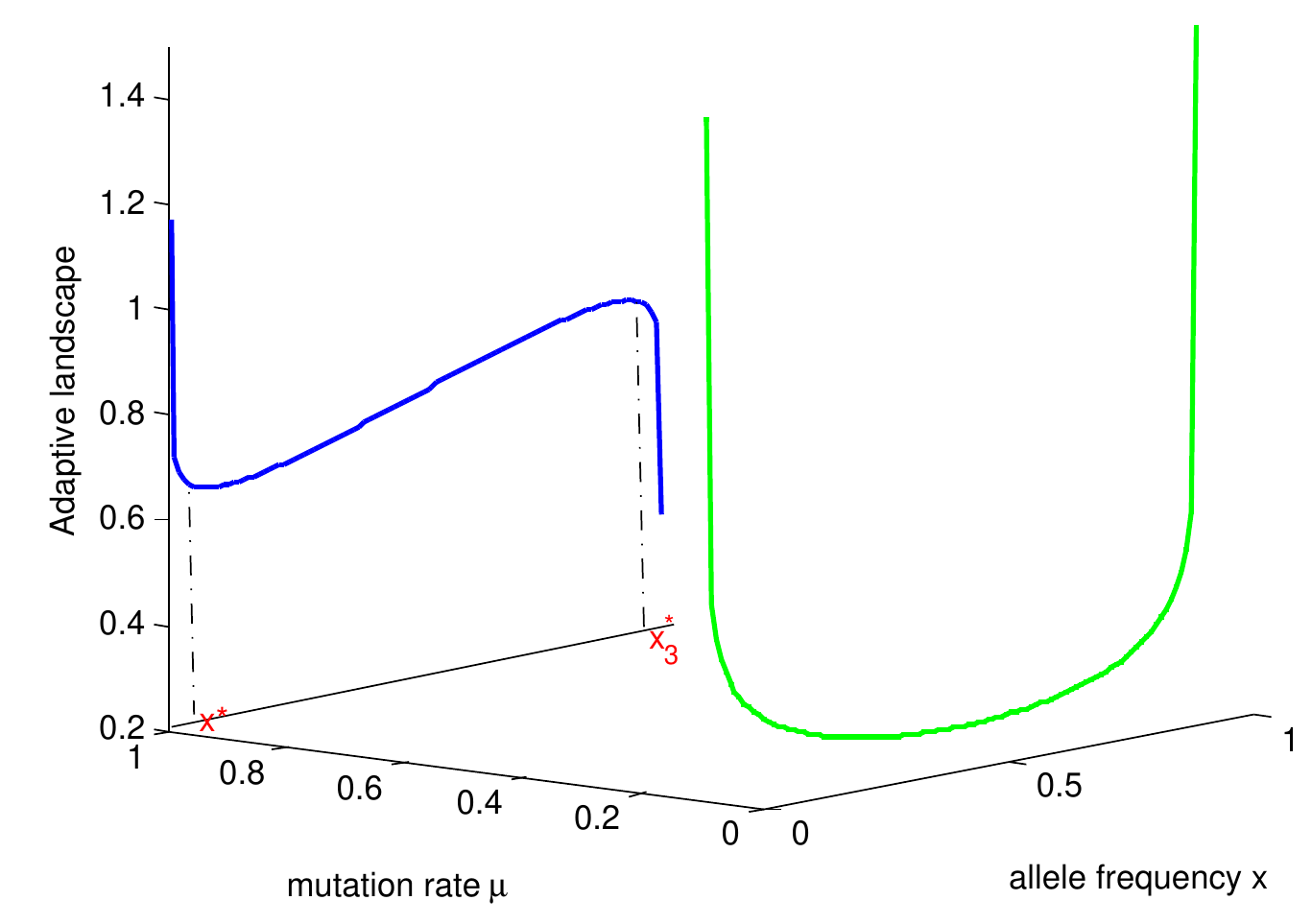}
\caption{\small \indent Adaptive landscape with two stable states in the low and high mutation rates. $N$=50, blue curve represents $\mu$=0.02, $\sigma$=0.1 while green curve stands for $\mu=0.000005$, $\sigma$=0.00005.}
\end{figure}
\indent  To evaluate the single click time and show the further power
of adaptive landscape, in the following we will demonstrate how the
single click time from one stable state to another is derived in this
framework. After straightforward calculation, backward Fokker- Planck
corresponding to Eq.(5) can be expressed with the property of time
homogeneous in the following form \cite{Kampen1992}\cite{Bokse2003}
\begin{equation}
\label{21}
\partial_{t}\rho=(f(x)+\epsilon D'(x))\partial_{x}\rho+\epsilon D(x)\partial_{x}^2\rho,
\end{equation}
 where the prime means the derivative of $D(x)$. General single click time $T(x)$ dependent on initial Dirac function satisfies
\begin{equation}
\label{22}
(f(x)+\epsilon D'(x))\partial_{x} T(x)+\epsilon D(x)\partial_{x}^2 T(x)=-1.
\end{equation}
With  boundary condition
\begin{equation}
T'(1)=0,\\
T(0)=0.
\end{equation}
Because of irreversible mutations from $A$ to $a$, the boundary is reflecting at $x=1$, the boundary  is absorbing at $x=0$ in the absence of mutation. \\
The general solution corresponding to Eq.(12) is
\begin{equation}
\label{25}
T(x)=\int_{x}^{0} \frac{1}{\epsilon D(y)} \exp (-\Phi(y))dy\int_{1}^{y} \exp(\Phi(z))dz,
\end{equation}
here $\Phi(x)=\int ^x f(x')/\epsilon D(x')dx' (\epsilon=1)$. \\
\indent Here the evolutionary process naturally occurs when
$x\in[0,1]$. In the process, there
are two important states $x^{\ast}$, $x_{3}^{\ast}$. Interval
$(0,1)$ contains an potential well at $x^{\ast}$,
$f(x^{\ast})=0$, $f'(x^{\ast})<0$, and an potential barrier at $x_{3}^{\ast}$,
$f(x_{3}^{\ast})=0$,
$f'(x_{3}^{\ast})>0$. Then one can use
asymptotic estimation to simplify Eq.(14) as
\begin{eqnarray}
T_{1\rightarrow 0}&=&\lim_{x\rightarrow 1}T(x) \nonumber \\
&\approx&\frac{2\pi \exp(\Phi(x_{3}^{\ast})-\Phi(x^{\ast}))}{D(x^{\ast})\sqrt{\Phi''(x^{\ast})\Phi''(x_{3}^{\ast})}} \nonumber \\
\quad &&+\int_{x_{3}^{\ast}}^{1}\frac{1}{f(x)}dx.
\end{eqnarray}
The second term is for downhill relaxation. The first term is  for barrier crossing.
This method for mean first passage time is consistent with \cite{qian2011}.\\
In the high mutation rates regime,  where $x_{3}^{\ast}$ approximates to a stable state which is near enough to $1$,
 $x^{\ast}$ corresponds to the saddle point that the population lies between the two stable states $0$ and $x_{3}^{\ast}$.  The single click time approximates to

\begin{eqnarray}
  \lefteqn{T_{1\rightarrow 0}=\lim_{x\rightarrow 1} T(x) }\nonumber \\
&&=  \lim_{x\rightarrow 1}\int_{0}^{x} \dfrac{1}{\epsilon D(y)}\exp (-\Phi(y))dy \int_{y}^ 1  \exp(\Phi(z))dz  \nonumber  \\
&&= \lim_{x\rightarrow 1}2N\int_{0}^{x}\dfrac{(1-y)^{\frac{2N\mu(\sigma-1)}{1-\sigma\mu}}}{(1-\sigma+y\sigma(1-\mu))^{\frac{2N(1-\mu)}{1-\sigma\mu}}}dy\nonumber\\
&& \quad\times\int_{y}^{1}z^{-1}(1-z)^{\frac{\sigma\mu-1-2N\mu(\sigma-1)}{1-\sigma\mu}}\nonumber\\
&& \quad \times(1-\sigma+\sigma z(1-\mu))^{\frac{2N(1-\mu)}{1-\sigma\mu}}dz\nonumber\\
&&\approx\frac{2\pi \exp(\Phi(x_{3}^{\ast})-\Phi(x^{\ast}))}{D(x^{\ast})\sqrt{\Phi''(x^{\ast})\Phi''(x_{3}^{\ast})}} \nonumber \\
&&\approx\dfrac{N}{3}2^{5-2N}\frac{(4-\sigma+3\sigma\mu)}{(2-\sigma\mu+3\sigma)}^{\frac{2N(1-\mu)}{1-\sigma\mu}}\mathcal{O}(1)
\end{eqnarray}
Here $x_{3}^{\ast}$ and $x^{\ast}$ are the zero points of $f(x)$ respectively. For the lower mutation rates regime, the single click time can be calculated similarly, where $x_{3}^{\ast}$ approximates to $1$, it can be derived by mean field method.
$x^{\ast}$ corresponds to the saddle point that the population lies at the lowest potential. $x^{\ast}$ is the zero point of $f(x)$.

Analogous to the derivation of $T_{1\rightarrow 0}$, we can calculate
\begin{eqnarray}
  \lefteqn{T_{0\rightarrow 1}} \nonumber\\
&& = \lim_{x\rightarrow 0}\int_{1}^{x} \dfrac{1}{\epsilon D(y)}\exp (-\Phi(y))dy \int_{y}^ 0  \exp(\Phi(z))dz  \nonumber  \\
&&= \lim_{x\rightarrow 0}2N\int_{1}^{x}\dfrac{(1-y)^{\frac{2N\mu(\sigma-1)}{1-\sigma\mu}}}{(1-\sigma+y\sigma(1-\mu))^{\frac{2N(1-\mu)}{1-\sigma\mu}}}dy\nonumber\\
&& \quad \times\int_{y}^{0}z^{-1}(1-z)^{\frac{\sigma\mu-1-2N\mu(\sigma-1)}{1-\sigma\mu}}\nonumber\\
&& \quad \times(1-\sigma+\sigma z(1-\mu))^{\frac{2N(1-\mu)}{1-\sigma\mu}}dz\nonumber \\
&& = \infty.
\end{eqnarray}
And $T_{0\rightarrow 1}$ goes to infinity as a polynomial way. Here the single click time asymptotically conforms to
exponential form according to Eq.(15),
though the shapes of adaptive landscape are different. And  once the population arrives at the stable state $x=0$, the population is absorbed as Eq.(17) describes. The accurate graph of single click time $T_{1\rightarrow0}$ is the following:\\
\begin{figure}[h]\centering
\includegraphics[width=7cm]{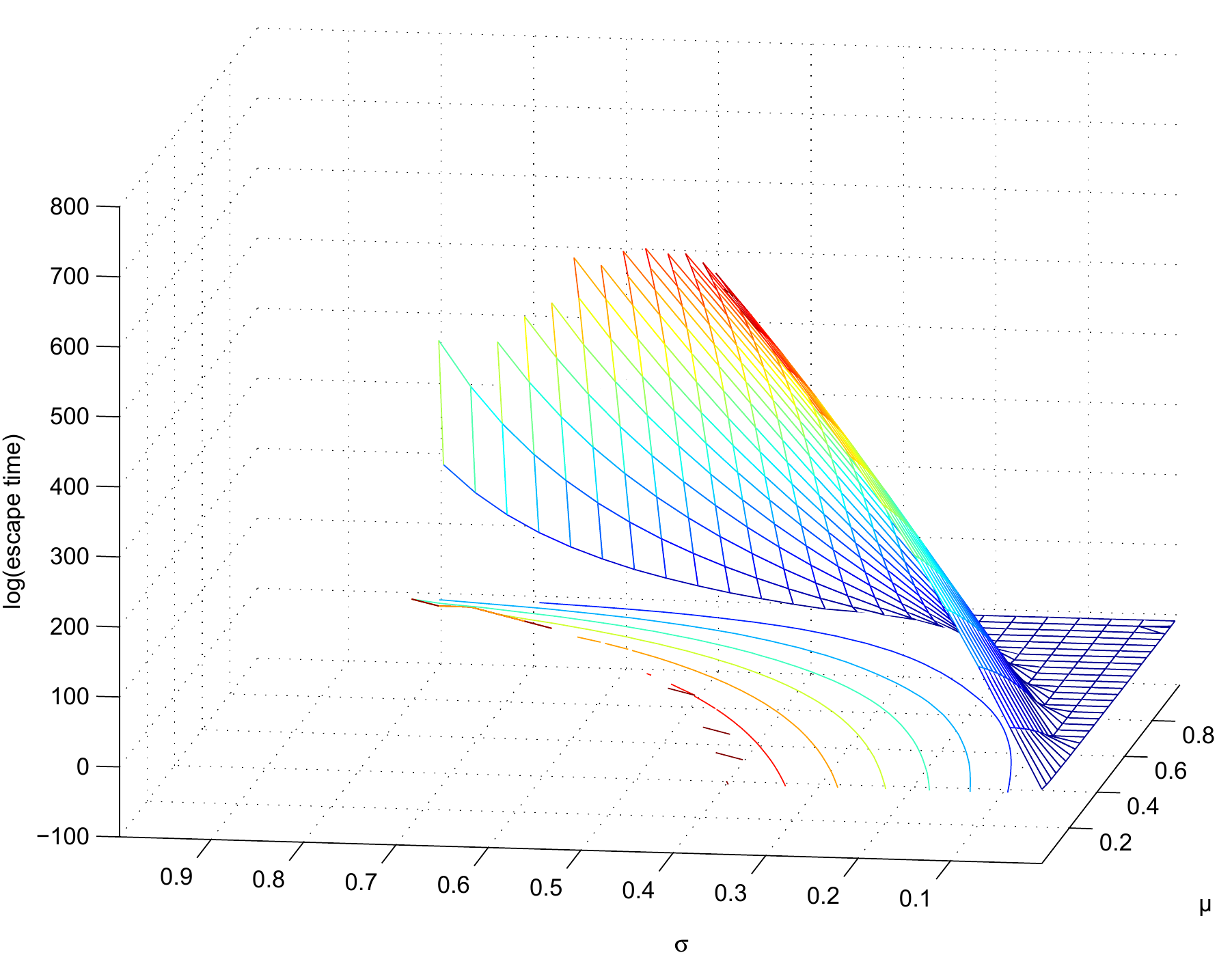}
\caption{\small \indent Exact numerical results about the single click time corresponding to Eq.(15). Horizontal coordinates represent selection rates $\sigma$ and mutation rates $\mu$, vertical coordinate corresponds to the  value of the click time $T_{1\rightarrow 0}$.}
\end{figure}
Fig.3 describes the single click time increases with increasing mutation rates and decreases with increasing selection rates. This is consistent with adaptive landscape in the lower and higher regimes in Fig.1. Asymptotic results corresponds to formula (16).

The estimated results  are more accurate when the parameters lie in the high mutation rates regime.
\subsection*{Discussion}
  The present article presents a different approach to estimate the single click time of Muller's ratchet. Previous works mainly based on the stochastic differential equations for diffusion approximation to study Muller's ratchet such as \cite{Gessler1995}\cite{Etheridge2009}. This method can result in the complex of calculating stochastic differential equations.
   Inspired by a recent work as \cite{waxman2010}, we connect it to one locus Wright-Fisher model with asexual population including $N$ haploid individuals.
   Direct classical Wright-Fisher model for diffusion approximation  can reduce complex calculation to solve  matrix equations, this method is functional especially when the dimension of the matrix is higher.
   Our theoretical results generalize the application of the model proposed in \cite{waxman2010}. And it does not need the existence and normalization of stationary distribution.  Our method investigates the global dynamical property of the system more directly. In addition, these results demonstrate  the single click time is approximately exponentially distributed. Most importantly, we depict the absorption phenomena from the adaptive landscape viewpoint. \\
  \indent To summarize, we have obtained two main sets of results in the
present work. First, we demonstrate a transformation of a simple
Fokker-Planck equation so that the adaptive landscape can be
explicitly read out as a potential function. Such demonstration
suggests this concept applicable to all levels of systems biology.
This not only allows computing click time of Muller's ratchet
straightforward, but also makes the stability analysis of the ratchet
system intuitive. Second, in this framework the derivation of single
click time is simple even for situations that steady state distribution
is nonnormalizable. Hence it allows us another way to handle absorbing
boundary condition. In this perspective our work may be a starting
point for estimating the click time for Muller's ratchet in more
general situations.
\section*{Acknowledgments}
\indent We would like to thank  Song Xu, Pengyao Jiang, Fangshu Cui, Quan Liu, Wei Zheng and others  in our lab for many conversations. This work was supported in part by the National 973
Projects No. 2007CB914700 and No. 2010CB529200
(P.A.)and in part by the project of Xinyang Normal university No. 20100073(S.J.).

\bibliographystyle{unsrt}
\bibliography{Muller}

\end{document}